\documentclass[a4paper]{article}

\usepackage{INTERSPEECH2019}
\usepackage{multirow}
\usepackage[table]{xcolor}
\usepackage{xcolor}

\title{Prosodic Prominence and Boundaries in Sequence-to-Sequence Speech Synthesis}
\name{ Antti Suni, Sofoklis Kakouros, Martti Vainio, Juraj \v{S}imko}

\address{Department of Digital Humanities, University of Helsinki, Finland}

\email{firstname.secondname@helsinki.fi}

\begin{document}

\maketitle
\begin{abstract}
Recent advances in deep learning methods have elevated synthetic speech quality to human level, and the field is now moving towards addressing prosodic variation in synthetic speech.
Despite successes in this effort, the state-of-the-art systems fall short of faithfully reproducing local prosodic events that give rise to, e.g.,  word-level emphasis and phrasal structure. This type of prosodic variation often reflects long-distance semantic relationships that are not accessible for end-to-end systems with a single sentence as their synthesis domain.  One of the possible solutions might be conditioning the synthesized speech by explicit prosodic labels, potentially generated using longer portions of text.

In this work we evaluate whether augmenting the textual input with such prosodic labels capturing word-level prominence and phrasal boundary strength can result in more accurate realization of sentence prosody. We use an automatic wavelet-based technique to extract such labels from speech material, and use them as an input to a tacotron-like synthesis system alongside textual information.

The results of objective evaluation of synthesized speech show that using the prosodic labels significantly improves the output in terms of faithfulness of $f_0$ and energy contours, in comparison with state-of-the-art implementations. 

  

\end{abstract}

\section{Introduction}

 

In the last few years, statistical speech synthesis has undergone a major paradigm shift.
Linguistic front-end (text normalization, letter-to-sound conversion, syllabification, part-of-speech tagging, phrasing, etc.) and acoustic back-end (source and filter parameters, deterministic vocoders, etc.) have been replaced by ``end-to-end'' systems, such as Tacotron and its derivatives \cite{wang2017tacotron,shen2018tacotron2,hayashi2019espnet}. Given a large enough corpus, the whole chain from raw text to speech (TTS) can now be jointly modelled with neural sequence-to-sequence (s2s) models, although the most successful applications in fact use elements of more traditional synthesis, e.g., separate text-normalization and text-to-phoneme conversion modules or separate (neural) vocoders.
Explicit modelling of prosodic parameters like segmental durations and pitch contours has been replaced by training attention mechanism and mapping textual input to acoustic properties represented with spectrograms. 

The s2s models, in particular when combined with WaveNet-style neural vocoders \cite{oord2016wavenet,shen2018tacotron2}, achieve quality on par with human speech, especially for isolated sentences with neutral prosody. In order to tackle more prosodically challenging tasks, 
neural architectures have been extended with techniques such as global style tokens \cite{wang2018style} and vector-quantized variational autoencoders \cite{van2017neural}. While these techniques yield impressive results in terms of modelling various global prosodic styles, they do not address prosodic variation on finer temporal scales, such as word-level emphasis and phrasal structure of the utterances. One of the reasons is the fact that this type of prosodic variation--for example emphasis associated with givenness of information--arises from long-distance dependencies in the text that falls between overall speech style and single sentence level prosody. 

Another, somewhat complementary reason arises from the lack of explicit control inherent to the ``black-box'' machine learning architectures, such as s2s systems. On the one hand, the existing systems are not designed to capture the long-range semantic dependencies \cite{clark2019evaluating}, on the other hand, they do not facilitate explicit control of prosody akin to older parametric synthesis approaches, where linguistic and prosodic labels were utilized and prosodic parameters were modelled separately \cite{zen2009statistical}.

In this paper we address the issue of explicit prosodic control by augmenting the textual material serving as an input to a s2s system with labels conveying word-level prominence and phrasal boundary strength. The labels are extracted from the training speech material using a wavelet-based technique \cite{suni2017hierarchical}. The objective evaluation shows that this type of local prominence and phrasal boundary control significantly contributes to the faithfulness of local prosodic variation in synthesized speech.


Prosodic labelling itself has a long history in TTS. (Binary) phrase break modelling has always been a necessary component of pre-s2s synthesis systems, and word prominence augmentation has also been experimented with, most recently in \cite{maliszpromis} where acoustically labeled prominence values were applied for controlling DNN-based parametric speech synthesis. Here, the achieved prosodic control was partly negated by a decrease in perceptual quality, likely due to parametric speech representation and a small database. In the s2s paradigm, pitch accent type have been used as an additional input for Japanese synthesis \cite{yasuda2019investigation}, demonstrating improvements in both subjective and objective evaluation, but the pitch accents were annotated manually. In contrast, the present study introduces automatic prosody annotation applied to s2s synthesis using a large training corpus.




\section{Methods and experiments}


\subsection{Prosodic Labelling}
\label{sec:CWToutline}

As the current synthesis models utilize tens of hours of training speech material, manual annotation of prosodic events is not a viable option. Instead, we thus use an automatic prominence annotation procedure providing word-level labels of acoustic prominence and boundary strength using a continuous wavelet transform (CWT) based method, described in full in \cite{suni2017hierarchical}. 

\begin{figure}[h!]
\centering
\includegraphics[width=1\linewidth]{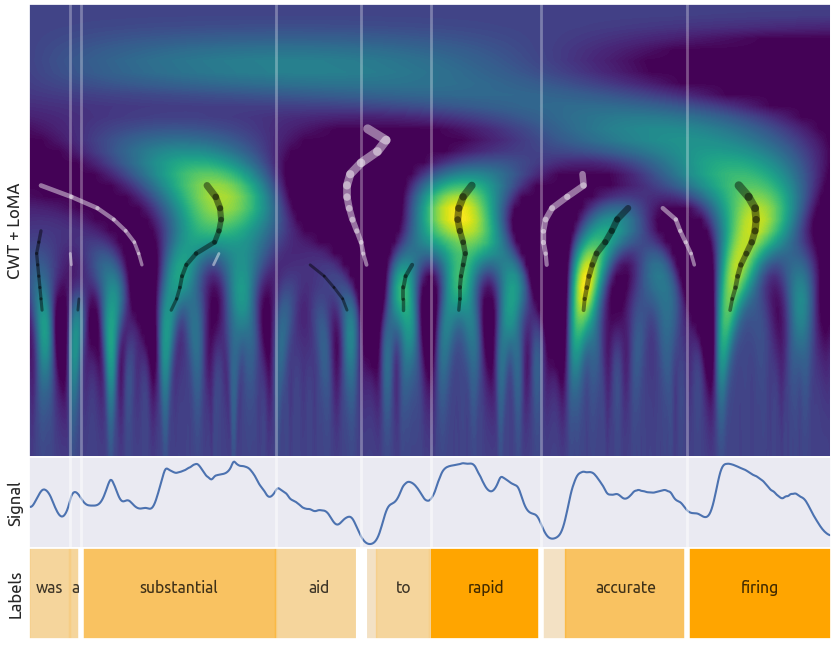}
\caption{CWT-based prosodic annotation method.}
\label{fig:cwt_pic}
\vspace{-1mm}
\end{figure}

The procedure first uses a forced-alignment of speech signal with the text. Subsequently, prosodic signals of $f_0$ and energy are extracted, and a word duration signal is created by placing the word duration value in a mid-point of each word and interpolating through the values. These three signals are then combined, and the combined signal  (signal panel in Figure~\ref{fig:cwt_pic}) is decomposed using CWT (heat map in Fig.~\ref{fig:cwt_pic}). 

CWT decomposes the signal into scales that can be associated with levels of prosodic hierarchy. To certain degree, events and movement related to, e.g., (prosodic) words and phrases can be isolated and analyzed, by following ridges or valleys across appropriate scales (black and white lines in Fig.~\ref{fig:cwt_pic}). Integrating the ridge / valley lines yields continuous word prominence / boundary estimates, that can be aligned with appropriate textual units. (The continuous word level prominence and boundary strength estimates and indicated by background saturation and thickness of word boundaries, respectively, in text panel of Fig.~\ref{fig:cwt_pic}). 

The method is essentially unsupervised, in that no labelled data are required, yet some degree of tuning is necessary regarding (language-dependent \cite{eriksson2018acoustic}) weights of the prosodic signal types as well as discretization of resulting continuous prominence and boundary values. For the current study, the weights were tuned according to the performance on an accent and boundary detection task on Boston radio news corpus \cite{ostendorf1995boston}, for which the method achieves state-of-the-art results \cite{suni2017hierarchical}. For word prominence estimates, the combined signal was calculated as a weighted sum of the $f_0$, energy and duration signals with weights 1.0, 0.5 and 1.0, respectively. For boundary strength estimates, three signals were multiplied. Both prominence and boundary values were discretized into three classes. The intervals were set manually, based on a small subset of the training utterances, such that for prominence, categories 0, 1 and 2 would roughly correspond to non-accented, accented and emphasized words. For boundaries, the phonological parallels would be no boundary, intermediate phrase boundary and intonational phrase boundary.  


\subsection{Sequence-to-sequence models}

Sequence-to-sequence models, like Tacotron \cite{wang2017tacotron} are trained to generate speech spectra directly from textual input in an end-to-end fashion, with prosodic features of represented speech learned jointly alongside spectral characteristics.
A front-end neural encoder encodes the textual input using a deep network combining convolutional and recurrent layers. A decoder (another stack of recurrent and convolutional layers) is trained to generate spectra in an auto-regressive, frame-by-frame manner; generation of each new frame is conditioned by a previously generated portion of the spectrogram. To provide dependency on text, the decoder is also conditioned by an output of an attention mechanism that time-aligns the output of the text encoder with the current state of the decoder network.

In the present evaluation of influence of prosodic labeling, we use a third party implementation \cite{KyubyongDCTTS} of tacotron-like architecture called Deep Convolutional TTS (DCTTS; \cite{tachibana2018dctts}). The DCTTS model replaces the recurrent layers used in the Tacotron system with dilated convolutions and highway layers, and uses a guided attention mechanism. These design decisions alleviate the high cost of training of recurrent layers and standard attention module, and the system trains considerably faster than the original Tacotron network, without loss in output quality. See \cite{tachibana2018dctts} for implementation and evaluation details of the system.

As in the original DCTTS implementation, the decoder module initially produces a downsampled coarse version of a MEL spectrogram that is subsequently upsampled using a Spectrogram Super-resolution Network (SSRN). We use the pre-trained SSRN from \cite{KyubyongDCTTS} for this purpose. Finally, the Griffin-Lim algorithm is used to generate an appropriate speech waveform from the full spectrogram \cite{griffin1984signal}.   


\subsection{Material}


The synthesis models were trained on a large, single-speaker American English corpus LJSpeech\cite{ljspeech17}, consisting of approximately 24 hours of non-fiction stories read by a  professional female reader. This dataset is commonly used in deep learning speech synthesis, due to its public availability, size, and consistent, if slightly reverberant quality.
Importantly for the current study, the reading style is quite expressive and the material consists of full chapters rather than isolated sentences. Informal listening reveals plenty of instances of long-range dependencies in prosody, in e.g. placing of contrastive or emphatic accents.

For prosodic labelling purposes, the dataset was aligned with Montreal forced aligner \cite{mcauliffe2017montreal}, using Librivox recipe. The discrete prosodic labels were then generated as described in Section~\ref{sec:CWToutline}, using authors' implementation of the process\footnote{https://github.com/asuni/wavelet\_prosody\_toolkit}. 
\subsection{Implementation}
To prepare the transcripts for training, the texts were phonemized (including stress marks) using CMU pronunication lexicon \cite{weide2014carnegie} with common punctuation (,.!?) included. Appropriate prosodic labels were simply inserted into text as additional symbols, with the word prominence labels preceding the words, and boundary labels following the words. For example, a text fragment \textit{`I insist, that'} would (if an emphasis was detected on \textit{insist} with a major boundary following it) be converted to:

\begin{scriptsize}\begin{verbatim} 
<p1> ay1 <b0> <p2> ih2 n s ih1 s t , <b2> <p0> dh ae1 t 
\end{verbatim}
\end{scriptsize}
The dataset transcriptions augmented with prosodic labels will be available online\footnote{https://www.mv.helsinki.fi/home/asuni/sp2020/}.

DCTTS s2s synthesis framework \cite{KyubyongDCTTS} (with the same architecture and hyper-parameters as in the original implementation) was used to train four models differing only in the prosodic marks used to augment the input: a baseline model without prosodic labels (\textbf{dctts}), a model with both prominence and boundary labels (\textbf{P+B}) and two models with either prominence or boundary labels (\textbf{P} and \textbf{B}). 

Of the 13,100 text fragments in the corpus 12,000 were used for training. Final chapter from the held-out data, 150 fragments were used for objective evaluation. The models were trained for 1000 epochs, and the test material was synthesized using oracle prosodic labels, i.e., the labels obtained for the test material in the same way as for the training set. For synthesis, the forcibly incremental attention procedure \cite{tachibana2018dctts} was relaxed, as the default implementation appeared to generate too fast speech. 

\section{Evaluation}

\begin{figure}
\centering
\includegraphics[width=1\linewidth]{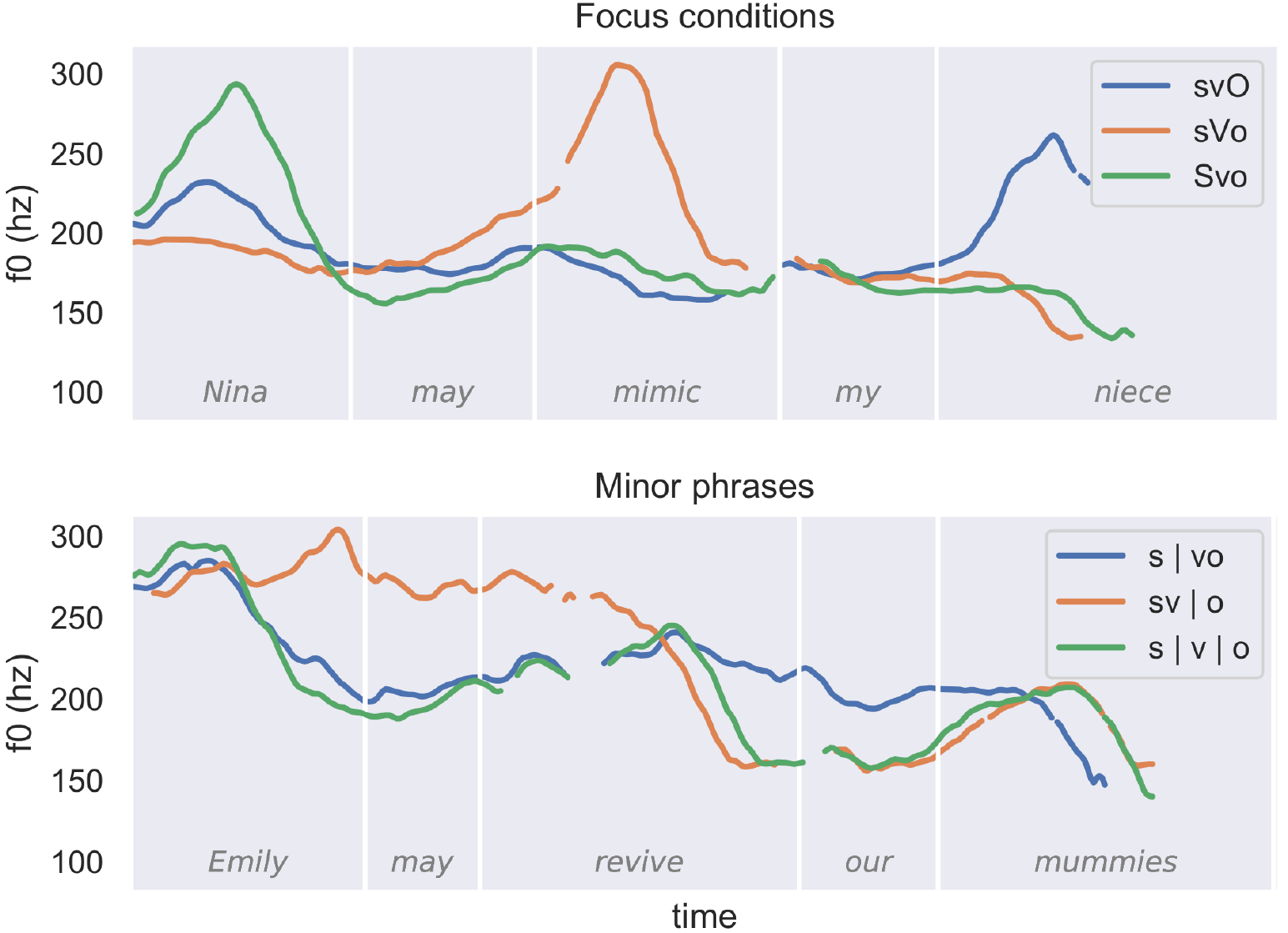}
\caption{Controlling focus placement and minor phrases.}
\label{fig:focus}
\vspace{-4mm}
\end{figure}

In order to evaluate the systems, we performed (1) a small qualitative assessment of prosody control, (2) a comparison between the baseline and the prosody-augmented systems, by re-labelling the prosody of synthetic test utterances and comparing the estimated labels to the reference labels, and (3) a standard objective evaluation by quantitatively comparing the prosodic signals extracted from synthetic utterances to those of the reference speech.  For the objective evaluation, we included two additional TTS baselines from state-of-the-art public framework\cite{hayashi2019espnet}, a Tacotron 2 (\textbf{taco}) and Tranformer (\textbf{trans}) model, which, with more complex model structure, could hypothetically model long-range dependencies of sentence prosody better than DCTTS framework. Note, that the prosody-augmented results do not reflect realistic TTS performance, but the ideal performance of such systems; instead of predicting the labels from text, we apply the oracle labels of reference speech.

\subsection{Assessment of Control}
First, we informally assessed the performance of the \textbf{P+B} model and the utility of the prosodic labels for control over prosodic realization of synthetic speech.

Short subject-verb-object sentences adapted from \cite{xu2005phonetic} were synthesized, simulating different focus and boundary conditions by manually setting the prosodic labels in the input. The top panel in Fig.~\ref{fig:focus} shows $f_0$ contours of three synthesized utterances with the emphasis location (elicited by the word prominence label \texttt{<p2>}) on subject, verb and object, respectively. Note that the system is able to reproduce the intended foci. In the bottom panel in Fig.~\ref{fig:focus}, intermediate, or minor phrases, boundaries were elicited by setting the boundary label to \texttt{<b1>} after subject, verb or both. Again, the three conditions are clearly differentiated, forming seemingly appropriate $f_0$ contours. 


\subsection{Categorical Evaluation}

\begin{table}
\small\addtolength{\tabcolsep}{-4pt}
\noindent\begin{tabular}{c|ccl|ccl}
 &  \bf dctts & & & \bf P + B & &  \\
\hline
  &  prec. & rec.  & F  & prec. & rec. &  F \\
\hline
  \bf prominence & & acc=0.61 & & & acc=0.81 &   \\
\hline
 \texttt{<p0>}    &   0.77   &   0.78   &   0.78 & 0.90   &   0.90  &    0.90 \\
 \texttt{<p1>}    &   0.35   &   0.44   &   0.39 & 0.61   &   0.59  &    0.60 \\
 \texttt{<p2>}    &   0.57   &   0.42   &   0.48 & 0.79   &   0.80  &    0.80 \\
\hline
 \bf boundary & & acc=0.70 & & & acc=0.85 & \\
\hline
 \texttt{<b0>}    &   0.78   &   0.82   &   0.80 & 0.88   &   0.92  &    0.90 \\
 \texttt{<b1>}    &   0.53   &   0.50   &   0.51 & 0.78   &   0.72  &    0.75 \\
 \texttt{<b2>}    &   0.52   &   0.43   &   0.47 & 0.86   &   0.84  &    0.85 \\

\end{tabular}
\vspace{2mm}
\caption{Synthetic vs original prosodic labels}
\vspace{-7mm}
    \label{tab:dataset}
  \end{table}

Test samples were synthesized by the baseline \textbf{dctts} system and the prosodically augmented \textbf{P+B} system.
For the latter, we used the oracle prosodic labels obtained from the original waveforms. The synthesised test samples were then prosodically labeled using the same procedure (Section~\ref{sec:CWToutline}), and the resulting labels were compared with the original ones.

Table~\ref{tab:dataset} summarizes the results of the comparison for the two systems. Accuracy, precision, recall, and the F-values are higher for the prosodically augmented \textbf{P+B} system than for the baseline.  The differences are greater for the emphatically accented (2) and the major boundary labels (2) than for to the unaccented and no-boundary labels (0 and 1), respectively. This indicates, that while the baseline system performed well above chance in producing distinguishable patterns of binary prominence and boundary in correct locations, it did not capture the finer distinctions of weak and strong prominence.


The \textbf{P+B} system struggled most in distinguishing the middle categories, which is expected due to somewhat arbitrary discretization of the inherently continuous prominence and boundary values.
 
 
 
\subsection{Objective Evaluation}

A set of objective measures comparing the original (reference) and the synthesized signals was used to formally evaluate the performance of the synthesis models. Although objective measures do not directly correlate with subjective measures of human perception, they provide the means to assess the overall model performance (see, e.g. \cite{wu2016investigating,Gu2018}). 

\begin{figure}[hb]
\centering
\includegraphics[width=1\linewidth]{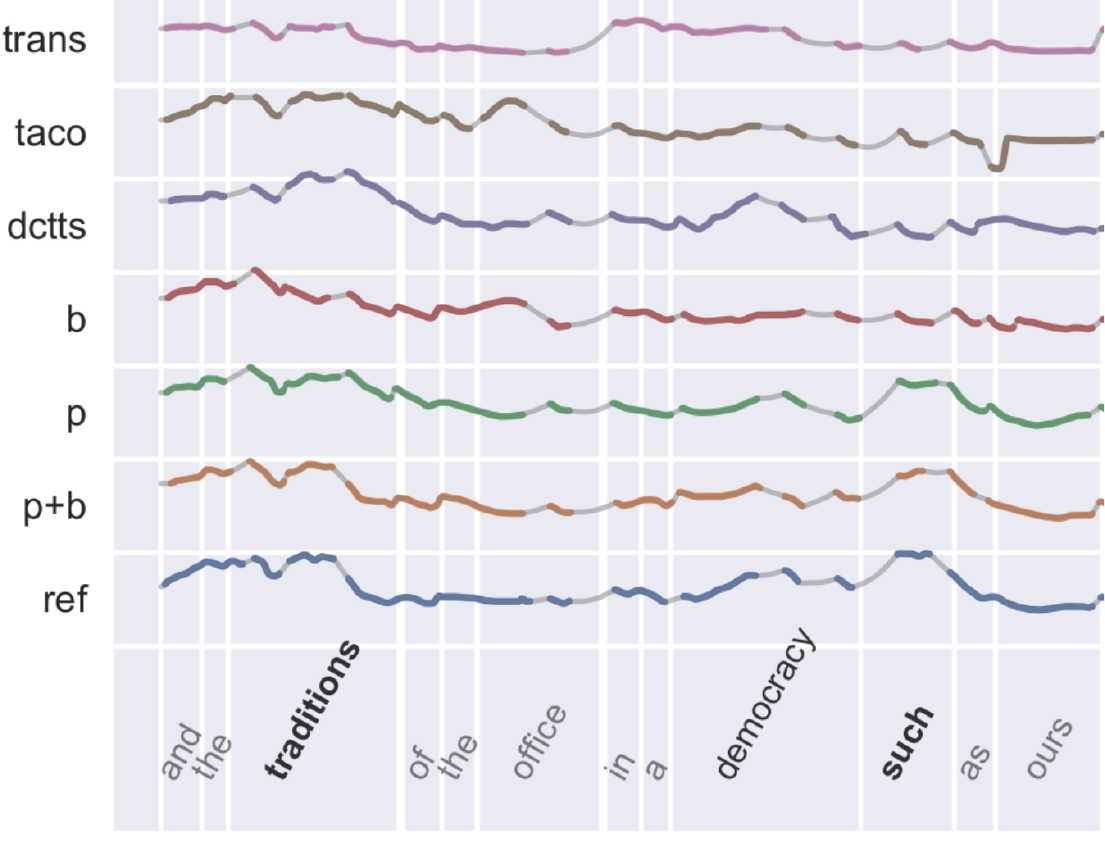}
  \caption{Comparison of methods in $f_0$ generation, see text.}
    \label{fig:comparison}
\end{figure}

The objective measures used in the current setup are the root-mean-squared error (RMSE) and Pearson correlation between the reference and the synthesized signal in terms of (i) the $f_0$ over the voiced intervals, (ii) the voiced energy, (iii) phone duration, and (iv) word duration.

To account for the mismatch in the time-alignment between the original and synthesized signals, the waveforms (original-synthesized) were compared using dynamic time warping (DTW) \cite{sakoe1978dynamic}, and the respective features were time-aligned to match the minimum distance score across the sequences. For the computation of the word and phone metrics, segmental and word durations were extracted by forced alignment using the Montreal Forced Aligner \cite{mcauliffe2017montreal}. Note that these steps (DTW, $f_0$ extraction and alignment) introduce some additional noise to the measurements. Fig.~\ref{fig:comparison} shows examples of $f_0$ contours of the utterances synthesized by the evaluated s2s systems, time aligned with respect to the original reference rendition (ref); the word-shading and boundary thickness reflect oracle labels.

\begin{table}[!bh]
\small\addtolength{\tabcolsep}{-2.5pt}
\noindent
\begin{tabular}{l|ccc|c|cc|l}
	&	{\scriptsize \textbf{P+B}}	&	{\scriptsize \textbf{P}}	&	{\scriptsize \textbf{B}}	&	{\scriptsize \textbf{dctts}}	&	{\scriptsize \textbf{taco}}	&	{\scriptsize \textbf{trans}} \\
	\hline
	\hline
$f_0$	&   \cellcolor[rgb]{0.6,0.6,0.6}2.132	&	\cellcolor[rgb]{0.6,0.6,0.6}2.339	&	\cellcolor[rgb]{0.8,0.8,0.8}2.507	&	2.639	&	2.954	&	2.865 & \parbox[t]{2mm}{\multirow{4}{*}{\rotatebox[origin=c]{270}{RMSE}}}\\ 
energy	&	\cellcolor[rgb]{0.6,0.6,0.6}3.245	&	\cellcolor[rgb]{0.6,0.6,0.6}3.240	&	3.315	&	3.642	&	3.456	&	3.359 & \\
ph.dur	&	0.025	&	0.025	&	0.026	&	0.027	&	0.027	&	0.029 &  \\
wd.dur	&	\cellcolor[rgb]{0.6,0.6,0.6}0.052	&	0.056	&	0.056	&	0.060	&	0.064	&	0.070 &  \\
\hline
\hline
$f_0$	&	\cellcolor[rgb]{0.6,0.6,0.6}0.655	&	\cellcolor[rgb]{0.6,0.6,0.6}0.595	&	0.519	&	0.471	&	0.460	&	0.457 &  \parbox[t]{2mm}{\multirow{4}{*}{\rotatebox[origin=c]{270}{Correl.}}}\\
energy	&	\cellcolor[rgb]{0.6,0.6,0.6}0.677	&	\cellcolor[rgb]{0.6,0.6,0.6}0.661	&	\cellcolor[rgb]{0.6,0.6,0.6}0.653	&	0.605	&	0.620	&	0.627 &  \\
ph.dur	&	\cellcolor[rgb]{0.6,0.6,0.6}0.833	&	\cellcolor[rgb]{0.8,0.8,0.8}0.825	&	\cellcolor[rgb]{0.8,0.8,0.8}0.825	&	0.798	&	0.788	&	0.770 &  \\
wd.dur	&	\cellcolor[rgb]{0.6,0.6,0.6}0.978	&	0.975	&	0.974	&	0.970	&	0.967	&	0.957 &  \\
\hline
\end{tabular}
\vspace{2mm}
\caption{RMSE and correlation values for the evaluated s2s systems for $f_0$, energy, phone duration, and word duration. Units: semitones for $f_0$, spl for energy,  and seconds for durations.}
\label{tab:objmeas}
\vspace{-4mm}
\end{table}


Table~\ref{tab:objmeas} lists the mean RMSE and correlation coefficients for the  values calculated for individual sentences in the test corpus, separately for comparisons between the reference and the utterances generated by different tested s2s systems. One-way anova (with a Bonferroni adjustment to compensate for multiple comparisons) was used to compare the values for different systems. The shaded cells in Table~\ref{tab:objmeas} indicate the attributes for which the prosody-augmented systems yielded significantly lower RMSE / higher correlation coefficient than the non-augmented system \textbf{dctts} (darker shade: $p<0.001$, lighter shading: $p<0.01$). As can be seen, the augmented systems performed significantly better for $f_0$ in terms of RMSE and correlation (except for \textbf{B} for the latter), and for energy (except RMSE for the \textbf{B} system). The prosody augmented systems also provide significantly higher correlations (but not lower RMSEs) between the aligned synthesized and reference signals in terms of phone duration. The system using both prosodic labels also reproduces the original word duration significantly better than the baseline.

\begin{figure}[t]
\centering
\includegraphics[width=\linewidth]{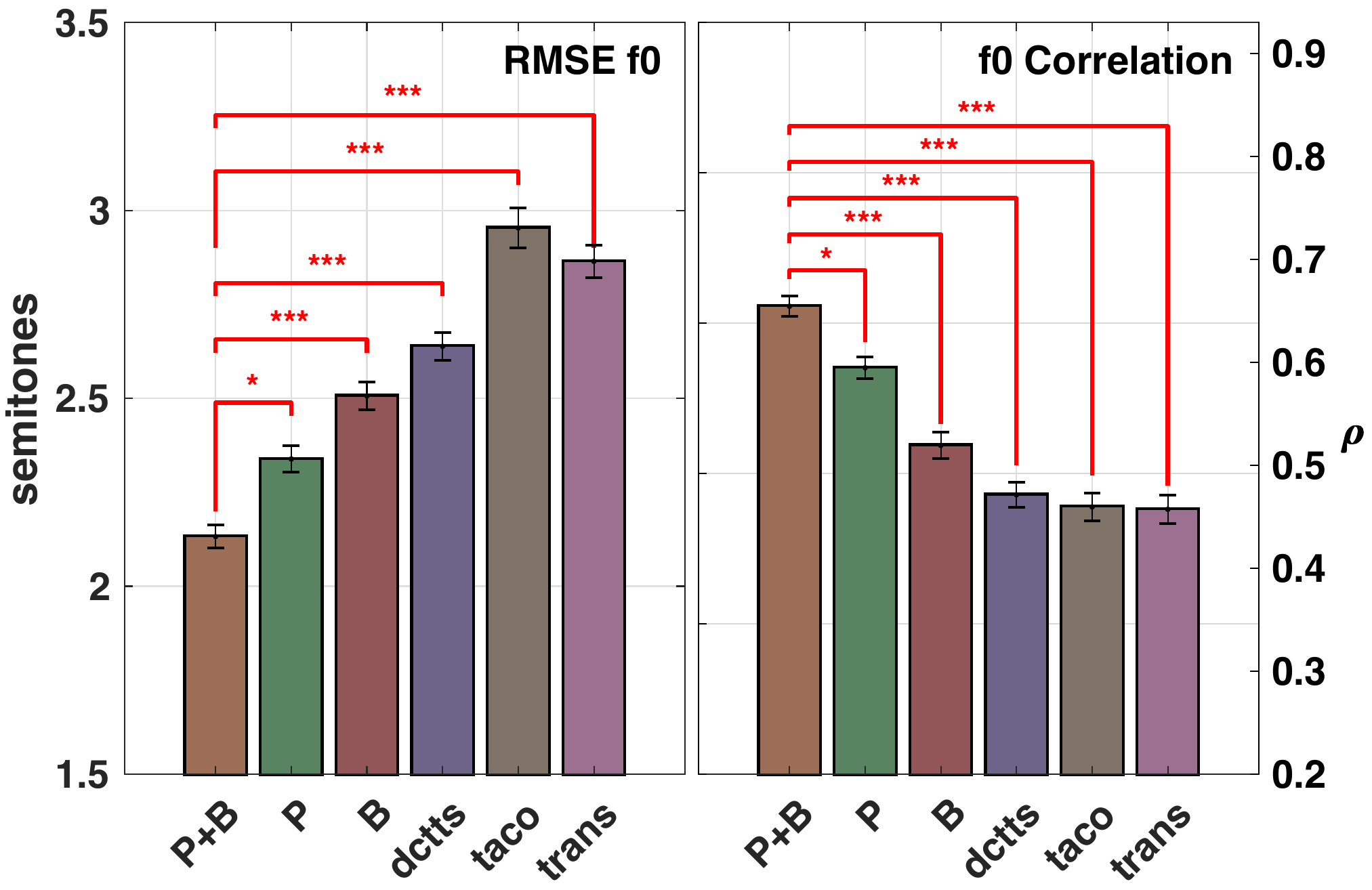}
  \caption{Mean RMSE and correlation coefficients for voiced $f_0$ for all tested systems (note the truncated y-axis). Significance of the difference between the \textbf{P+B} system and all other systems is indicated in red. }
    \label{fig:rmse_corr_f0}
    \vspace{-5mm}
\end{figure}

For the great majority of the assessed measures, the \textbf{P+B} achieved the lowest mean RMSEs and the highest mean correlations of all evaluated systems. This performance advantage is particularly strong for $f_0$ measure. As shown in Fig.~\ref{fig:rmse_corr_f0}, for $f_0$ \textbf{P+B} in fact performs significantly better than any other of the evaluated systems.
In general, both prominence \textbf{P} and boundary \textbf{B} labels improve upon baseline, and the effects are independent of each other, as combining both label types yields further improvements in most measures. Neither \textbf{taco} nor \textbf{trans} TTS model yield improvements upon \textbf{dctts} baseline, in fact their matching of the reference $f_0$ is quite poor, despite high perceptual quality of those systems \cite{hayashi2019espnet}.

%




\section{Discussion}

The current study shows, that prosodic control and reproduction of prosodic patterns of natural speech is to a high degree achievable within a sequence-to-sequence synthesis paradigm. The next important step will be an evaluation of how do the 
the measurable improvements translate to perceptual quality, in particular for longer stretches of textual input.

In this work we use oracle prosodic labels extracted from the target speech material. In order to develop a fully fledged TTS system, these prosodic labels need to be predicted directly from the text.
A recent contribution \cite{talman2019predicting} has yielded promising results in predicting prosodic labels such as those used here from a text using contextualized word representations that can capture long-term semantic dependencies in text. 

While the systems using prosodic labels reproduce signal characteristics better than the TTS baseline, the correlations with the reference speech for the prosody augmented system are still relatively low (e.g., 0.655 for $f_0$, etc.). It should be noted that our labeling scheme operates on slow scales of words and phrases rather than on syllables used as the fundamental unit for many prosody annotation schemes such as ToBI. While it might be unreasonable to expect that an s2s system could learn to perfectly handle local phenomena such as accent type, peak-alignment and boundary tones, the presented approach would in principle allow for a shift of the scope of prominence labels to syllable level. 

Striving for such level of descriptive adequacy could, however be excessive: enforcement of too much detail could, in our opinion, hamper the ability of s2s system to generalize from the training material. Also, considerations of what can be robustly labeled from acoustics and what can be predicted from text must also be taken into account. Hence, we believe that word and phrase level labeling forms a good trade-off. 







\section{Acknowledgements}

The work was partly funded by an Academy of Finland Research Fellowship grant (\#309575) to the last author.

\newpage

\bibliographystyle{IEEEtran}

\bibliography{mybib}

\begin{thebibliography}{10}
\providecommand{\url}[1]{#1}
\csname url@samestyle\endcsname
\providecommand{\newblock}{\relax}
\providecommand{\bibinfo}[2]{#2}
\providecommand{\BIBentrySTDinterwordspacing}{\spaceskip=0pt\relax}
\providecommand{\BIBentryALTinterwordstretchfactor}{4}
\providecommand{\BIBentryALTinterwordspacing}{\spaceskip=\fontdimen2\font plus
\BIBentryALTinterwordstretchfactor\fontdimen3\font minus
  \fontdimen4\font\relax}
\providecommand{\BIBforeignlanguage}[2]{{%
\expandafter\ifx\csname l@#1\endcsname\relax
\typeout{** WARNING: IEEEtran.bst: No hyphenation pattern has been}%
\typeout{** loaded for the language `#1'. Using the pattern for}%
\typeout{** the default language instead.}%
\else
\language=\csname l@#1\endcsname
\fi
#2}}
\providecommand{\BIBdecl}{\relax}
\BIBdecl

\bibitem{wang2017tacotron}
Y.~Wang, R.~Skerry-Ryan, D.~Stanton, Y.~Wu, R.~J. Weiss, N.~Jaitly, Z.~Yang,
  Y.~Xiao, Z.~Chen, S.~Bengio \emph{et~al.}, ``Tacotron: Towards end-to-end
  speech synthesis,'' \emph{arXiv preprint arXiv:1703.10135}, 2017.

\bibitem{shen2018tacotron2}
J.~Shen, R.~Pang, R.~J. Weiss, M.~Schuster, N.~Jaitly, Z.~Yang, Z.~Chen,
  Y.~Zhang, Y.~Wang, R.~Skerrv-Ryan \emph{et~al.}, ``Natural tts synthesis by
  conditioning wavenet on mel spectrogram predictions,'' in \emph{2018 IEEE
  International Conference on Acoustics, Speech and Signal Processing
  (ICASSP)}.\hskip 1em plus 0.5em minus 0.4em\relax IEEE, 2018, pp. 4779--4783.

\bibitem{hayashi2019espnet}
T.~Hayashi, R.~Yamamoto, K.~Inoue, T.~Yoshimura, S.~Watanabe, T.~Toda,
  K.~Takeda, Y.~Zhang, and X.~Tan, ``Espnet-tts: Unified, reproducible, and
  integratable open source end-to-end text-to-speech toolkit,'' \emph{arXiv
  preprint arXiv:1910.10909}, 2019.

\bibitem{oord2016wavenet}
A.~v.~d. Oord, S.~Dieleman, H.~Zen, K.~Simonyan, O.~Vinyals, A.~Graves,
  N.~Kalchbrenner, A.~Senior, and K.~Kavukcuoglu, ``Wavenet: A generative model
  for raw audio,'' \emph{arXiv preprint arXiv:1609.03499}, 2016.

\bibitem{wang2018style}
Y.~Wang, D.~Stanton, Y.~Zhang, R.~Skerry-Ryan, E.~Battenberg, J.~Shor, Y.~Xiao,
  F.~Ren, Y.~Jia, and R.~A. Saurous, ``Style tokens: {U}nsupervised style
  modeling, control and transfer in end-to-end speech synthesis,'' 2018.

\bibitem{van2017neural}
A.~van~den Oord, O.~Vinyals \emph{et~al.}, ``Neural discrete representation
  learning,'' in \emph{Advances in Neural Information Processing Systems},
  2017, pp. 6306--6315.

\bibitem{clark2019evaluating}
R.~Clark, H.~Silen, T.~Kenter, and R.~Leith, ``Evaluating long-form
  text-to-speech: Comparing the ratings of sentences and paragraphs,''
  \emph{arXiv preprint arXiv:1909.03965}, 2019.

\bibitem{zen2009statistical}
H.~Zen, K.~Tokuda, and A.~W. Black, ``Statistical parametric speech
  synthesis,'' \emph{speech communication}, vol.~51, no.~11, pp. 1039--1064,
  2009.

\bibitem{suni2017hierarchical}
A.~Suni, J.~{\v{S}}imko, D.~Aalto, and M.~Vainio, ``Hierarchical representation
  and estimation of prosody using continuous wavelet transform,''
  \emph{Computer Speech \& Language}, vol.~45, pp. 123--136, 2017.

\bibitem{maliszpromis}
Z.~Malisz, H.~Berthelsen, J.~Beskow, and J.~Gustafson, ``Promis: a
  statistical-parametric speech synthesis system with prominence control via a
  prominence network,'' in \emph{Proc. 10th ISCA Speech Synthesis Workshop},
  2019, pp. 257--262.

\bibitem{yasuda2019investigation}
Y.~Yasuda, X.~Wang, S.~Takaki, and J.~Yamagishi, ``Investigation of enhanced
  tacotron text-to-speech synthesis systems with self-attention for pitch
  accent language,'' in \emph{ICASSP 2019-2019 IEEE International Conference on
  Acoustics, Speech and Signal Processing (ICASSP)}.\hskip 1em plus 0.5em minus
  0.4em\relax IEEE, 2019, pp. 6905--6909.

\bibitem{eriksson2018acoustic}
A.~Eriksson, A.~S. Suni, M.~T. Vainio, J.~Simko \emph{et~al.}, ``The acoustic
  basis of lexical stress perception,'' in \emph{Proceedings of the 9th
  International Conference on Speech Prosody 2018}.\hskip 1em plus 0.5em minus
  0.4em\relax International Speech Communications Association, 2018.

\bibitem{ostendorf1995boston}
M.~Ostendorf, P.~J. Price, and S.~Shattuck-Hufnagel, ``The {B}oston
  {U}niversity radio news corpus,'' \emph{Linguistic Data Consortium}, pp.
  1--19, 1995.

\bibitem{KyubyongDCTTS}
K.~Park. (2018) {A TensorFlow Implementation of DC-TTS: yet another
  text-to-speech model}. \url{https://github.com/Kyubyong/dc\_tts}. Accessed:
  2019-09-29.

\bibitem{tachibana2018dctts}
H.~Tachibana, K.~Uenoyama, and S.~Aihara, ``Efficiently trainable
  text-to-speech system based on deep convolutional networks with guided
  attention,'' in \emph{2018 IEEE International Conference on Acoustics, Speech
  and Signal Processing (ICASSP)}.\hskip 1em plus 0.5em minus 0.4em\relax IEEE,
  2018, pp. 4784--4788.

\bibitem{griffin1984signal}
D.~Griffin and J.~Lim, ``Signal estimation from modified short-time fourier
  transform,'' \emph{IEEE Transactions on Acoustics, Speech, and Signal
  Processing}, vol.~32, no.~2, pp. 236--243, 1984.

\bibitem{ljspeech17}
K.~Ito, ``The lj speech dataset,''
  \url{https://keithito.com/LJ-Speech-Dataset/}, 2017.

\bibitem{mcauliffe2017montreal}
M.~McAuliffe, M.~Socolof, S.~Mihuc, M.~Wagner, and M.~Sonderegger, ``{M}ontreal
  {F}orced {A}ligner: Trainable text-speech alignment using kaldi.'' in
  \emph{Interspeech}, 2017, pp. 498--502.

\bibitem{weide2014carnegie}
R.~Weide, ``The carnegie mellon pronouncing dictionary [cmudict. 0.7b],'' 2014.

\bibitem{xu2005phonetic}
Y.~Xu and C.~X. Xu, ``Phonetic realization of focus in english declarative
  intonation,'' \emph{Journal of Phonetics}, vol.~33, no.~2, pp. 159--197,
  2005.

\bibitem{wu2016investigating}
Z.~Wu and S.~King, ``Investigating gated recurrent networks for speech
  synthesis,'' in \emph{2016 IEEE International Conference on Acoustics, Speech
  and Signal Processing (ICASSP)}.\hskip 1em plus 0.5em minus 0.4em\relax IEEE,
  2016, pp. 5140--5144.

\bibitem{Gu2018}
\BIBentryALTinterwordspacing
Y.~Gu and Y.~Kang, ``Multi-task wavenet: A multi-task generative model for
  statistical parametric speech synthesis without fundamental frequency
  conditions,'' in \emph{Proc. Interspeech 2018}, 2018, pp. 2007--2011.
  [Online]. Available: \url{http://dx.doi.org/10.21437/Interspeech.2018-1506}
\BIBentrySTDinterwordspacing

\bibitem{sakoe1978dynamic}
H.~Sakoe and S.~Chiba, ``Dynamic programming algorithm optimization for spoken
  word recognition,'' \emph{IEEE transactions on acoustics, speech, and signal
  processing}, vol.~26, no.~1, pp. 43--49, 1978.

\bibitem{talman2019predicting}
A.~Talman, A.~Suni, H.~Celikkanat, S.~Kakouros, J.~Tiedemann, and M.~Vainio,
  ``Predicting prosodic prominence from text with pre-trained contextualized
  word representations,'' \emph{arXiv preprint arXiv:1908.02262}, 2019.

\end{thebibliography}


\end{document}